\documentclass[a4paper,11pt]{article}
\pdfoutput=1 % if your are submitting a pdflatex (i.e. if you have
\usepackage{jheppub} % for details on the use of the package, please
\usepackage[T1]{fontenc} % if needed

\title{\boldmath Anisotropic Scale Invariant Spacetimes and Black Holes in Zwei-Dreibein Gravity}

\author{A. F. Goya}
\affiliation{Universidad de Buenos Aires FCEN-UBA and IFIBA-CONICET,\\Ciudad Universitaria, Pabell\'on I, 1428, Buenos Aires, Argentina}
\emailAdd{af.goya@df.uba.ar}

\abstract{
We show that Zwei-Dreibein Gravity (ZDG), a bigravity theory recently proposed by Bergshoeff, de Haan, Hohm, Merbis, and Townsend in ref. \cite{ZDG}, admits exact solutions with anisotropic scale invariance. These type of geometries are the three-dimensional analogues of the spacetimes which were proposed as gravity duals for condensed matter systems. In particular, we find Schr\"{o}dinger invariant spaces as well as Lifshitz spaces with arbitrary dynamical exponent $z$. We also find black holes that are asymptotically Lifshitz with $z=3$, showing that these (non-constant curvature) solutions of New Massive Gravity (NMG) are persistent after the introduction of the infinite tower of higher-curvature terms of ZDG, provided a renormalization of the parameters. Black holes in asymptotically warped Anti-de Sitter spaces are also found. Interestingly, in almost all the geometries studied in this work, the metric associated with the second dreibein turns out to be equivalent, up to a constant global factor, to the first one. This phenomenon has been previously observed in other bigravity theories in asymptotically flat and asymptotically Anti-de Sitter backgrounds. However, for the particular case of the $z=3$ Lifshitz black hole, here we found that the second metric corresponds to a different black hole that coincides with the former only in the asymptotic region. In fact, we find a new family of $z=3$ black holes that corresponds to a one-parameter deformation of the NMG solution.
}

\begin{document} 
\maketitle
\flushbottom

\section{Introduction}
\label{Sec:Intro}
The idea that a theory of quantum gravity must be `holographic', in the sense that the relevant degrees of freedom must be `on the surface', was surrounding the theoretical physics community for several years \cite{HolographicPrinciple1,HolographicPrinciple2}. The first concise and most successful realization of holographic ideas was given by Maldacena in 1997 \cite{Maldacena:AdSCFT}, relating type IIB string theory in five-dimensional Anti-de Sitter space (AdS) with $\mathcal{N} = 4$ Super Yang-Mills theory in four-dimensional Minkowski space. Since then, it is commonly believed that this gravity/gauge duality \cite{Witten:AdSCFT,GKP98} is an intrinsic property of any sensible candidate for quantum gravity, so it must hold not only for string theory but for any gravity theory that is expected to be well defined beyond the classical limit. Therefore, this relation must establish a connection between any complete gravity theory on AdS$_{d+1}$ and a conformal field theory (CFT) in $d$-dimensional Minkowski space.

Since 2008, there have been attempts to extend AdS/CFT in order to apply holography to non-relativistic field theories scenarios, like superconductors \cite{HolographicSC} and other problems in condensed matter \cite{HolographicCMTLectures}. These proposals to generalize AdS/CFT had even involved non-AdS bulk geometries; such is the case of spacetimes having the same symmetries as the Schrödinger group with anisotropic scale invariance, which were conjectured to be gravity duals of strongly coupled condensed matter systems \cite{HolographicSchroedinger,NonRelatDuals}; Other example is the so-called Lifshitz spacetimes, which were proposed as gravity duals for Lifshitz fixed points \cite{LifshitzDual}. The question then arised as to what kind of gravity theory would admit these anisotropic scale invariant spaces as exact solutions. Notably, a three-dimensional higher-derivative gravity theory, dubbed as New Massive Gravity (NMG) \cite{NMG,MoreNMG}, was shown to contain in its spectrum both Schr\"{o}dinger spaces \cite{NMG:AdSWaves} and Lifshitz spaces \cite{NMG:LifshitzBH} as exact solutions. In addition, the theory also contains Lifshitz black holes \cite{NMG:LifshitzBH}, which would be dual to finite temperature systems of these type. This invites to consider NMG in the context of condensed matter applications of holography, at least as a toy model. On the other hand, NMG also admits as exact solutions other interesting non-AdS spaces that could be relevant for holography, such as Warped Anti-de Sitter (WAdS) geometries, which are streched or squashed deformation of AdS$_3$, and which could provide another interesting playground to study extensions of the AdS/CFT correspondence. The isometry group of WAdS$_3$ is $SL(2,\mathbb{R}) \times U(1)$ and provides enough information to build a boundary field theory \cite{WCFT}. This kind of spacetimes and the black holes that asymptote to them are solutions of NMG as well \cite{Clement:2009gq}. All this convinces ourselves that three-dimensional massive gravity seems to be a good arena to study different non-AdS extensions of holography.

Very recently, a new 3D gravity model with very interesting properties, called Zwei-Dreibein Gravity (ZDG), was proposed \cite{ZDG}. It was shown in \cite{ZDG:LogAdSWaves} that this bigravity theory can be viewed as a higher-derivative gravity theory of which NMG is a particular limit. Some indications were found in ref. \cite{ZDG:LogAdSWaves} that ZDG at some particular points of the parameter space can be interpreted as a gravity dual of the Logarithmic Conformal Field Theories (LCFT) \cite{LCFT}, which are also relevant in the context of condensed matter. Hence, a natural question is whether ZDG can also be used in non-relativistic extensions of the AdS/CFT correspondence. In order to answer this we should verify whether Lifshitz, Schr\"{o}dinger and Warped AdS spacetimes are solutions of ZDG as well. This would give a hint that ZDG could provide another sensible model to test non-relativistic holography.

In this paper we will show that, indeed, the recently proposed ZDG theory does admit all those spaces as exact solutions. We will construct explicit solutions of ZDG representing Schr\"{o}dinger spacetimes as well as black holes that asymptote to both Lifshitz and WAdS spacetimes. An interesting feature observed in other bigravity theories considered in the literature is that, when a black hole solution is considered, the two `metrics' of the theory are related each other by coordinate transformations and a global rescaling. Here, we will show that this sort of `duality' also appears in ZDG for some solutions. However, for the asymptotically Lifshitz black hole, the second `metric' represents a new Lifshitz black hole that coincides with the first one only in the asymptotic region. In fact, here we find a new one-parameter family of Lifshitz black holes.

The paper is organized as follows: In section \ref{Sec:ReviewNMG}, we will review NMG theory and its non-AdS solutions. In sections \ref{Sec:ZDG} and \ref{Sec:ZDGSolutions}, we will present ZDG theory and explicitly show how the non-AdS geometries of NMG persists in this three-dimensional bigravity theory, and we will give the explicit relation between the two spin-two fields. Finally in section \ref{Sec:Conclusions}, we will conclude with some final remarks and outlook.

\section{Review of New Massive Gravity}
\label{Sec:ReviewNMG}
In this section, we will review the important aspects of New Massive Gravity (NMG) \cite{NMG,MoreNMG} and the relevant solutions of this theory for this work.

NMG is a higher-derivative gravity theory that at linearized level is equivalent to a Fierz-Pauli theory for a massive spin 2 field. Its action is given by the Einstein-Hilbert term (with a sign parameter $\varsigma = \pm 1$), a cosmological constant term, and a particular term quadratic in the curvature; namely
\begin{equation*}
\label{NMG Action}
	S_{NMG} = \dfrac{M_P}{2} \int d^3x \sqrt{-g} \left( \varsigma R - 2\Lambda + \dfrac{1}{m^2} \left(R_{\mu\nu} R^{\mu\nu} - \dfrac{3}{8}R^2 \right) \right) \,,
\end{equation*}
where the free parameters of the theory are the Planck mass $M_P$, the bare cosmological constant $\Lambda$, and the mass of the gravitons $m^2$. The specific choice of the relative coefficient $-3/8$ between the squared curvature terms is necessary in order to eliminate a scalar ghost mode from the spectrum.

An interesting feature of NMG is that it contains a large variety of solutions. In particular, due to its higher-derivative nature, it is possible to circumvent the obstructions of Birkhoff-like theorems and find different static spherically symmetric solutions. The spectrum of NMG contains hairy black holes, wormholes, gravitational solitons, and kinks \cite{Oliva:2009ip}. It is also possible to find asymptotically AdS$_3$ spaces with weaker boundary conditions relative to the Brown-Henneaux ones \cite{Giribet:2010ed,Giribet:2011vv}. Remarkably, this theory also supports asymptotically dS$_3$ black holes \cite{MoreNMG,Oliva:2009ip}, unlike three-dimensional General Relativity. Besides, as we said before, NMG provides a playground for non-relativistic holography since it contains solutions whose isometries realizes non-relativistic symmetries. Let us review some of these solutions below.

\subsection{Lifshitz Spacetimes in NMG}
\label{SubSec:NMGLifshitz}
Lifshitz spacetimes have an anisotropic scale invariance under
\begin{equation*}
t \rightarrow \lambda^z t \,, \quad x \rightarrow \lambda x \,.
\end{equation*}
As we said before, geometries that realize this kind of symmetries were conjectured to be gravity duals of condensed matter systems of Lifshitz type \cite{LifshitzDual}. The so-called Lifshitz spacetimes in $d+1$ dimensions are given by
\begin{equation}
	ds^2 = -\dfrac{r^{2z}}{\ell^{2z}}dt^2 + \dfrac{\ell^2}{r^2}dr^2 + \dfrac{r^2}{\ell^2}d\vec{x} \cdot d\vec{x} \,,
	\label{Lifshitz}
\end{equation}
where $\vec{x}$ is a $(d-1)$-dimensional spacelike vector. It is worth noticing that for $z=1$ the spacetime (\ref{Lifshitz}) correspond to AdS$_{d+1}$. The three-dimensional version of (\ref{Lifshitz}) is an exact solution of NMG \cite{NMG:LifshitzBH}. The curvature radius $\ell$ and the mass parameter $m^2$ are related with the parameter $z$ by
\begin{align}
\label{Lifshitz NMG Cond }
m^2 \ell^2  = \dfrac{1}{2\varsigma} (z^2 - 3z + 1) \,, \quad \Lambda \ell^2 = -\dfrac{\varsigma}{2} (z^2 + z + 1) \,.
\end{align}

Notably, NMG also contains asymptotically Lifshitz black holes as we will see below.

\subsection{Lifshitz Black Hole in NMG}
\label{NMG:LifshitzBH}
In order to study condensed matter theories at finite temperature it is necessary to find black holes that coincide with the Lifshitz space in the asymptotic region. In \cite{NMG:LifshitzBH} a black hole was found that is asymptotically Lifshitz with $z=3$. It was recently shown that this is the only asymptotatically Lifshitz black hole with $z \neq 1$ in the theory \cite{LifshitzBHUniqueness}\footnote{At least for spacetimes that can be written as a Kerr-Schild deformation of a Lifshitz space.}. The metric of this three-dimensional $z=3$ Lifshitz black hole is given by the following line element
\begin{equation*}
ds^2 = -\dfrac{r^6}{\ell^6}\left(1-\dfrac{M\ell^2}{r^2}\right) dt^2 + \dfrac{r^2}{\ell^2} dx^2 + \left(\dfrac{r^2}{\ell^2}-M\right)^{-1}dr^2  \,.
\end{equation*}

The metric above solves the equations of motion of NMG provided
\begin{equation*}
	m^2 \ell^2  = \dfrac{1}{2\varsigma} \,, \quad \Lambda \ell^2 = -\dfrac{13\varsigma}{2} \,.
\end{equation*}

The above conditions are just (\ref{Lifshitz NMG Cond }) evaluated at $z=3$. We will see that this solution also persists in Zwei-Dreibein Gravity (ZDG) \cite{ZDG}.

\subsection{Schr\"{o}dinger Spacetimes in NMG}
\label{SubSec:NMGSchroedinger}
Other interesting geometries for non-AdS holography applications are Schr\"{o}dinger spaces. These spaces realize geometrically the non-relativistic conformal group, also known as the Schr\"{o}dinger group. These spacetimes are particular cases of AdS waves \cite{NMG:AdSWaves} and are also equivalent to null WAdS spaces, which we will discuss in the next subsection.
The Schr\"{o}dinger line element is given by
\begin{equation*}
	ds^2 = \dfrac{\ell^2}{y^2} \left( -\dfrac{\ell^{2\nu}}{y^{2\nu}}du^2 -2dudv + dy^2 \right) \,,
\end{equation*}
and happens to be a solution of NMG if
\begin{equation*}
	m^2\ell^2 = -2\varsigma\big(1+8\nu(\nu+1)\big) \,, \quad \Lambda = \dfrac{1-4m^2\ell^2\varsigma}{4m^4\ell^4} \,.
\end{equation*}

Next, let us discuss the WAdS spaces.

\subsection{Warped Anti-de Sitter Spacetimes in NMG}
\label{SubSec:NMGWAdS}
Warped Anti-de Sitter space can be obtained as a Hopf fibration of $\mathbb{R}$ over AdS$_2$ and multiplying the fiber by a constant warp factor. In a convenient system of coordinates the line element reads
\begin{equation*}
	ds^2 = \frac14\left(-\cosh^2 x \, d\tau^2 + dx^2 + \dfrac{4\nu^2}{\nu^2+3}(dy+\sinh x \, d\tau)^2 \right) \,.
\end{equation*}

For $\nu=1$ this line element is equivalent to AdS$_3$ in a rotating frame. There are different types of deformations of AdS$_3$ producing timelike, spacelike, and null WAdS$_3$ spaces. As we said before, the Schr\"{o}dinger space correspond to the null WAdS$_3$. It is interesting to notice that the timelike version of WAdS$_3$ actually corresponds to the three-dimensional part of the G\"{o}del solution of General Relativity \cite{Goedel}\footnote{The author thanks Geoffrey Comp\`{e}re and St\'{e}phane Detournay for clarifying comments about this subject.}. The G\"{o}del spacetime can be written as a direct product $\Sigma_3 \times \mathbb{R}$, and this $\Sigma_3$ piece is precisely the timelike WAdS$_3$ geometry. In this work we will focused on the spacelike WAdS$_3$. Notably, there are exact solutions that represent black holes that are locally and asymptotically equivalent to spacelike WAdS$_3$ spaces in the same way as how the BTZ black hole is locally and asymptotically equivalent to AdS$_3$. The WAdS$_3$ black holes are given by
\begin{subequations}
\label{WAdSADM}
\begin{equation}
ds^{2}=-N^{2}dt^{2}+\rho ^{2}\left( d\varphi +N^{\varphi }dt\right) ^{2}+
\dfrac{\ell^{2}dr^{2}}{4\rho ^{2}N^{2}} \,,
\label{dsWAdS}
\end{equation}
with 
\begin{equation}
\begin{split}
\rho ^{2} &=\dfrac{r}{4}\left(3(\nu ^{2}-1)r+(\nu ^{2}+3)(r_{+}+r_{-})- 4\nu \sqrt{(\nu ^{2}+3) r_{+}r_{-}}\right), \\
N^{2} &=\dfrac{(\nu ^{2}+3)(r-r_{+})(r-r_{-})}{4\rho ^{2}}, \\
N^{\varphi } &=\dfrac{2\nu r-\sqrt{(\nu ^{2}+3)r_{+}r_{-}}}{2\rho ^{2}} \,.
\end{split}
\label{WAdSadm}
\end{equation}
\end{subequations}

In the context of parity even massive gravity theories, these black holes were first discussed by Cl\'{e}ment in ref. \cite{Clement:2009gq}. They are solutions of NMG if $m^2$ and $\Lambda$ satisfy the following relations
\begin{equation*}
	m^2\ell^2 = -\dfrac{(20\nu^2-3)\varsigma}{2} \,, \quad \Lambda\ell^2 = \dfrac{\varsigma (9-48\nu^2+4\nu^4)}{2(20\nu^2-3)} \,. \nonumber
\end{equation*}

The WAdS$_3$ black holes of NMG were studied in detail in ref. \cite{WAdSBHTonni}; see also \cite{WAdSBHGoya}.

\section{Zwei-Dreibein Gravity}
\label{Sec:ZDG}
Now, let us describe the main aspects of the Zwei-Dreiben Gravity theory (ZDG) proposed in ref. \cite{ZDG}.

ZDG is a bigravity theory whose dynamical fields are the two dreibein 1-forms $e_{I}{}^a = e_I{}_{\mu}{}^adx^{\mu}$ and the two spin-connection 1-forms $\omega_{I}{}^a = \omega_I{}_{\mu}{}^adx^{\mu}$. While the subindex $I= 1,2$ identifies each of the two spin-2 fields, the superindex $a=\hat{1},\hat{2},\hat{3}$ refers to the coordinates in the tangent bundle. Index $\mu = 0,1,2$ labels spacetime coordinates. The Lagrangian of the theory is given by
\begin{equation}
\begin{split}
	L_{ZDG} =  - \mathcal{M}_P \bigg\{ & \sigma e_{1\,a} R_1{}^a +  e_{2\,a} R_2{}^a +  \frac{1}{6}m^2 \epsilon_{abc} \left(\alpha_1 e_1{}^a e_1{}^b   e_1{}^b   
  +  \alpha_2 m^2  e_2{}^a e_2{}^b  e_2{}^c \right) \\
 &  - \frac{1}{2}m^2 \epsilon_{abc} \left( \beta_1 e_1{}^a e_1{}^b e_2{}^c 
 + \beta_2 e_1{}^a e_2{}^b e_2{}^c \right)  \bigg\} \, ,
\end{split}
\label{Lbimetric}
\end{equation}
where we suppressed the wedge symbol $\wedge$ and where all the products between forms must be understood as exterior products.

The curvature and the torsion 2-forms are given by the following expressions
\begin{equation*}
R_I{}^a = {\rm d} \omega_I{}^a + \frac12 \epsilon^{abc} \omega_{I\,b} \omega_{I\,c}\, , \quad
T_I{}^a = \mathcal{D}_I e_I{}^a
\equiv {\rm d} e_I{}^a + \epsilon^{abc} \omega_{I\,b} e_{I\,c} \,,
\end{equation*}
where the usual rank-2 valued 2-forms have been replaced by its dual tensors, e.g. $\mathcal{T}^a = (1/2) \epsilon^a_{\:\: b c} \: \mathcal{T}^{b c}$.

The independent parameters in (\ref{Lbimetric}) are the two cosmological constant parameters $\alpha_I$, the two interaction constants $\beta_I$, and the Planck mass $\mathcal{M}_P$. The parameter $m^2$ is a convenient but redundant mass parameter, and $\sigma = \pm 1$ is a sign parameter.

For arbitrary values of the parameters $\beta_1$ and $\beta_2$ the theory contains three bulk degrees of freedom, two corresponding to the helicity $\pm 2$ modes of a massive graviton. The remaining degree of freedom is potentially a ghost mode. One way to avoid this presumably ghost mode is by restricting the parameter space of the theory, for example, choosing $\beta_2=0$, and assuming that $e_1$ is invertible \cite{Deffayet:2012nr,ZDG,Banados:2013fda,3DCSlikeHamiltonian}.\footnote{In order to remove the unwanted degrees of freedom from the spectrum, it is necessary to generate secondary constraints to eliminate them. One way to do that is to assume the invertibility of $\beta_1 e_1{}^a + \beta_2 e_2{}^a$. This requirement will guarantee extra secondary constraints. The choice made in this work ($e_1$ invertible and $\beta_2$=0) is a particular case of the former one. For more details see \cite{3DCSlikeHamiltonian}.}

Therefore, the equations of motion for $e_{1}{}^a, e_{2}{}^a$, and $\omega_{I}{}^a$, with $\beta_2=0$, and with an invertible $e_1$, derived from the Lagrangian density \eqref{Lbimetric} are given by:
\begin{subequations}
\label{Reom}
\begin{align}
\label{R1eom}
0  &= \;  \sigma R_1^a + \frac12 m^2 \epsilon^a_{\:\: bc} \left(
 \alpha_1 e_1^{\,b}e_1^{\,c} - 2 \beta_1 e_1^{\,b}e_2^{\,c} \right)\,,
 \\
\label{R2eom}
0  &=  \;  R_2^a + \frac12 m^2  \epsilon^a_{\:\: bc} \left(
\alpha_2 e_2^{\,b}e_2^{\,c} - \beta_1  e_1^{\,b} e_1^{\,c}  \right)\,,
 \\
\label{Teom}
0  &= \; T_{I}^a\,,
\end{align}
\end{subequations}
respectively.

Notice that the curvature and torsion 2-forms satisfy the Bianchi identities
\begin{equation*}\label{Bianchi}
\mathcal{D}_I R_I{}^a = 0\,, \quad 
\mathcal{D}_I T_I{}^a = \epsilon^{abc} R_{I\,b} e_{I\,c} \,.
\end{equation*}

The kinetic terms of $e_1$ and $e_2$ in \eqref{Lbimetric} are invariant under their respective diffeomorphisms and local Lorentz transformations. Notice that the presence of the interaction term breaks the symmetries to their diagonal subgroup, given by the identification of the gauge parameters of each group.

An interesting remark is that ZDG can be understood as a theory with a single dreibein that, however, contains an infinite number of higher-derivative terms \cite{ZDG:LogAdSWaves}. The fact that bigravity theories can be viewed as higher-derivative theories of one metric field was first noticed in \cite{Hassan:2013pca}. A similar procedure was followed in \cite{ZDG:LogAdSWaves} for the ZDG case. To see this, first we observe that we can solve the equations of motion (\ref{R1eom}) to obtain an expression for $e_2{}^{a}$ in terms of $e_1{}^{a}$. Then, using the property $\varepsilon^{\rho \sigma \tau} R_{1\, \sigma \tau}{}^a =  \det({e_1}) e_1{}^{\sigma\, a} G_1{}^{\rho}{}_{\sigma}$, we obtain the following expression for $e_2{}^{a}$:
\begin{equation}
	e_{2\, \mu}{}^{a} = \dfrac{\alpha_1}{2\beta_1} e_{1\, \mu}{}^{a} + \dfrac{\sigma}{m^2 \beta_1} S_{1\,\mu}{}^a \,,
	\label{e2}
\end{equation}
where $ S_{1\,\mu}{}^a \equiv S_{1\, \mu\nu} e_{1}^{\nu\, a}$ and where $S_{1\, \mu\nu} = R_{1\,\mu\nu} - \frac14 R_1 g_{1\,\mu\nu}$ is the Schouten tensor associated with the metric  $g_{1\,\mu\nu} \equiv e_{1\,\mu}{}^a e_{1\,\nu}{}^b \eta_{ab}$. Here, we identify $g_{1\,\mu\nu}$ with the physical metric since, as we said before, we have assumed $e_1$ to be invertible. The `auxiliary field' $e_2$ represents the higher-derivative content of the theory. The dreibein $e_2$ is auxiliary in the sense that it can be solved for algebraically from the equations of motion \eqref{R1eom}, in the same way as in the second derivative formulation of NMG \cite{Hohm:2010jc}.

It is worth pointing out that the resulting higher-derivative equations of motion for $g_1$ up to order $1/m^2$ can be integrated to an action if the parameters of the theory satisfy a relation consistent with the NMG limit (see Appendix B of \cite{ZDG:LogAdSWaves} for more details). This does not mean that ZDG with generic values of its coupling constant is a higher-derivative completion of NMG. However, we remark that the resulting action at order $1/m^2$ is precisely the NMG action. In general, the intrepretation of ZDG as a higher-derivative gravity theory of one metric, only holds at the level of the equations of motion without matter fields.

\subsection{A Relation Between $g_1{}_{\mu\nu}$ and $g_2{}_{\mu\nu}$}
\label{SubSec:SelfDuality}
Besides the relation \eqref{e2} between $e^{\mu}_{1\, a}$ and $e^{\mu}_{2\, a}$ and the corresponding metrics $g_{1\, \mu\nu} = e_{1\ \mu}^{\ a} e_{1\ \nu}^{\ b} \eta_{ab}$ and $g_{2\, \mu\nu} = e_{2\ \mu}^{\ a} e_{2\ \nu}^{\ b} \eta_{ab}$, one can wonder whether for a particular class of solutions, there exists a coordinate transformation that maps $g_2$ into $g_1$ and viceversa\footnote{The author thanks Gast\'on Giribet for draw his attention to this point, specially to the reference \cite{3DfgbigravityBH}.}. It is worth mentioning that here and in what follows we will refer to the $g_2$ field as a `metric', as a consequence of the form that it takes that resembles a metric. Since we are only assuming the invertibility of $e_1$, $g_1$ is the physical metric\footnote{In the footnote 4 of ref. \cite{ZDG:LogAdSWaves} the authors said that it is possible to find the inverse of $e_2$ as a power expansion in $1/m^2$. On the other hand, even when the invertibility on the dreibein $e_2$ is not imposed as a necessary condition, notice that the explicit solutions we found here for the two spin-two fields are manifestly invertible.}. Whether both spin-two fields have a clear geometric interpretation is not yet understood.

In this sub-section we will explore whether both 'metrics' $g_1$ and $g_2$ are related by coordinate/scale transformations. This sort of `duality' between the two spin-two fields were found in the so-called \textit{f-g theory} \cite{Isham:1971gm}, where the two metrics ($g_{\mu\nu}$ and $f_{\mu\nu}$) represent the Schwarzschild-(A)dS solution at the same time \cite{fgtheorysphsolutions}. In \cite{BiGravityHairyBH}, hairy black hole solutions were found that asymptote (A)dS in the Hassan-Rosen bigravity theory \cite{Hassan:2011zd} and a similar phenomenon happens there. All these are examples of four dimensional bigravity theories; however, there also exists an example of this type in three dimensions: In \cite{3DfgbigravityBH} the three dimensional version of the \textit{f-g theory} was studied and it was found asymptotically AdS$_3$ black hole solutions for the two metrics. Therefore, it is natural to ask whether ZDG has this `duality' as well. In the particular case in which $g_1$ is a maximally symmetric space, provided $R_1{}_{\mu\nu} = 2\Lambda g_1{}_{\mu\nu}$, using the expression \eqref{e2} it becomes clear that $g_2$ will be proportional to $g_1$. This is also true for the Ba\~{n}ados-Teitelboim-Zanelli (BTZ) black hole \cite{BTZ}, since it is locally equivalent to AdS$_3$. More precisely, if the metric associated to $e_1$ is given by 
\begin{equation*}
	ds_1{}^2 = -\left( \dfrac{r^2}{\ell^2} - M \right) dt^2 + 2J dt d\varphi + r^2d\varphi^2
	+ \left( \dfrac{r^2}{\ell^2} - M + \dfrac{J}{4r^2} \right)^{-1} dr^2 \,,
\end{equation*}
then, the `metric' associated to $e_2$ is given by a constant conformal factor times the BTZ black hole metric 
\begin{equation*}
\begin{split}
	ds_2{}^2 = \dfrac{1}{4m^2\ell^4\beta_1^2} \bigg\{ & -\left( \dfrac{r^2}{\ell^2} - M \right) dt^2 + 2J dt d\varphi + r^2d\varphi^2 + \\
	& \left( \dfrac{r^2}{\ell^2} - M + \dfrac{J}{4r^2} \right)^{-1} dr^2 \bigg\} \,.
\end{split}
\end{equation*}

The same happens for the AdS waves studied in \cite{ZDG:LogAdSWaves}. In the latter case, the metric associated to $e_1$ is given by
\begin{equation*}
	ds_1{}^2 = \dfrac{\ell^2}{y^2} \left\lbrace -f(u,y) du^2 - 2 du dv + dy^2 \right\rbrace \,,
\end{equation*}
where the wave profile $f(u,y)$ is polynomial in $y$
\begin{equation}
	f(u,y) = f_0(u) + f_2(u) \left( \dfrac{y}{\ell} \right)^2 + f_{+} \left( \dfrac{y}{\ell} \right)^{n_+} + f_{-}(u) \left( \dfrac{y}{\ell} \right)^{n_-} \,,
	\label{AdSWaveProfile}
\end{equation}
and $\left\lbrace 0,2,n_+,n_-\right\rbrace$ are the roots of the characteristic polynomial obtained from the equations of motion \eqref{R1eom}-\eqref{R2eom}\footnote{The special cases where the roots $n_{\pm}$ degenerate where studied in detail in \cite{NMG:AdSWaves} for the NMG case, and in \cite{ZDG:LogAdSWaves} for the ZDG case.}.

On the other hand, the `metric' of $e_2$ is given by
\begin{equation*}
	ds_2{}^2 = \dfrac{(m^2\ell^2\alpha_1-\sigma)^2}{4m^2\ell^4\beta_1^2} \dfrac{\ell^2}{y^2} \left\lbrace -F(u,y) du^2 - 2 du dv + dy^2 \right\rbrace \,,
\end{equation*}
where 
\begin{equation*}
	F(u,y) = \left( 1 - 2\sigma y ( y \partial_y^2 - \partial_y ) \right) f(u,y) \,,
\end{equation*}
is identified with the $g_2$ wave profile. Here we will assume that $m^2\ell^2 \alpha_1 \neq \sigma$.

Therefore, we observe that in ZDG we have a similar phenomenon: For some solutions, spin-two fields $g_1$ and $g_2$ represent the same type of geometry, what we mean is that when one dreibein field solution corresponds to a given geometry, the other dreibein field soltion is given by the same geometry up to parameter redefinitions and diffeomorphisms. It would be interesting to find out if the same `duality' holds in the case of non-AdS geometries. As far as we known, all the known solutions for which such a `duality' between metrics holds, either in three or four spacetime dimensions, are asymptotically flat, dS or AdS geometries. Here we will show that, indeed, ZDG bigravity theory, possesses non-AdS solutions which exhibit this sort of `duality' between the two dynamical fields. This suggests that this phenomenon is much more universal of what we thought and it is not necessarily related to the simplicity of asymptotically constant curvature solutions.

\section{New Exact Solutions of Zwei-Dreibein Gravity}
\label{Sec:ZDGSolutions}
In this section we will construct explicit non-AdS exact solutions of ZDG.

\subsection{Lifshitz Spacetimes in ZDG}
\label{SubSec:ZDGLifshitz}
Let us begin with Lifshitz type solutions. The expression (\ref{e2}) gives us a hint to propose an ansatz to find a Lifshitz solution of ZDG. It is convenient to rescale the parameters $\alpha_1$, $\alpha_2$, and $\beta_1$ by defining $\alpha \equiv m^2\ell^2 \alpha_1$, $\mathcal{A} \equiv m^2\ell^2 \alpha_2$, and $\beta \equiv m^2\ell^2 \beta_1$. The dreibein $e_1$ will be related to the Lifshitz metric in the usual way $g_1{}_{\mu\nu} = e_1{}_{\mu}{}^a\eta_{ab}e_1{}_{\nu}{}^b$; and using (\ref{e2}) the ansatz for $e_2{}^a$ is
\begin{equation}
\begin{split}
	e_2{}^{\hat{1}} &= \dfrac{ \alpha - \sigma (z^2+z-1)}{2\beta }  \left(\dfrac{r}{\ell}\right)^z dt \, , \\
	e_2{}^{\hat{2}} &= \dfrac{ \alpha + \sigma (z^2-z-1) }{ 2\beta } \left(\dfrac{r}{\ell}\right) dx \, , \\
	e_2{}^{\hat{3}} &= \dfrac{ \alpha - \sigma (z^2-z+1) }{ 2\beta } \left(\dfrac{\ell}{r}\right) dr \, .
\end{split}
\label{e2Lifshitz}
\end{equation}

Three-dimensional Lifshitz spacetime is a solution of the ZDG equations of motion \eqref{R1eom}-\eqref{R2eom} provided $\mathcal{A}$ and $\beta$ satisfy the following relations
\begin{equation}
\begin{split}
	\mathcal{A} &= -\dfrac{ \big( \alpha (\alpha - 2\sigma z ) - (z^2+z-1)(z^2-z-1) \big)^2 \big( \alpha - \sigma(z^2+z+1) \big) }{ 2\sigma \big( \alpha - \sigma(z^2-z+1) \big)^4 } \, , \\
	\beta &= - \dfrac{ \alpha ( \alpha - 2\sigma z ) - (z^2+z-1)(z^2-z-1) }{2\sigma \big( \alpha - \sigma(z^2-z+1) \big) } \,,
\end{split}
\end{equation}
or, equivalently,
\begin{equation}
	\dfrac{\mathcal{A}}{\beta} = \dfrac{ \big( \alpha ( \alpha - 2z\sigma) - (z^2+z-1)(z^2-z-1) \big) \, \left( \alpha - \sigma(z^2+z+1) \right) } { \big( \alpha - \sigma (z^2-z+1) \big)^3 } \,,
	\label{ZDG:LifshitzCond}
\end{equation}
assuming in particular $\alpha\sigma \neq z^2-z+1 $.

As we said in the previous section, it would be interesting to investigate whether there exists non-AdS solutions to bigravity theories like ZDG that exhibit the mentioned `duality' between the two spin-two fields. To give a first answer to this question we can observe that the `metric' associated with \eqref{e2Lifshitz} can be put in the following form
\begin{equation*}
	ds_2{}^2 = \Omega^2 \left( -\dfrac{r^{2z}}{\ell^{2z}}d\tilde{t}^{\:2} + \dfrac{r^2}{\ell^2} d\tilde{x}^2 + \dfrac{\ell^2}{r^2} dr^2 \right) \,,
\end{equation*}
where we have (re-)defined
\begin{equation*}
\begin{split}
	\Omega^2 &= \dfrac{ \left(\alpha - \sigma(z^2-z+1)\right)^2 }{4\beta^2} \,, 
	\quad \tilde{t} = \dfrac{ \alpha - \sigma(z^2+z-1) }{ \alpha - \sigma(z^2-z+1) } \,, \\
	\tilde{x} &= \dfrac{ \alpha - \sigma(z^2-z-1) }{ \alpha - \sigma(z^2-z+1) } \,.
\end{split}
\end{equation*}
Thus, $g_2$ also represents a Lifshitz spacetime with the same dynamical exponent $z$, that is a global rescaling of $g_1$, assuming additionally $\alpha \neq \sigma(z^2 \pm z-1)$.

\subsection{Lifshitz Black Holes in ZDG}
\label{SubSec:ZDGLifshitzBH}
Remarkably, for the special case $z=3$, ZDG also admits a static circularly symmetric black hole solution that asymptotes Lifshitz space.  This shows that the Lifshitz black hole of NMG
\begin{equation}
ds^2 = -\dfrac{r^6}{\ell^6}\left(1-\dfrac{M\ell^2}{r^2}\right) dt^2 + \dfrac{r^2}{\ell^2} dx^2 + \left(\dfrac{r^2}{\ell^2}-M\right)^{-1} dr^2  \,,
\label{LifthitzBH}
\end{equation}
persists as an exact solution after the infinite set of higher-curvature terms of ZDG were introduced\footnote{As we said in section \ref{Sec:ZDG}, ZDG is not a higher-curvature completion of NMG.}. This is notable due to the fact that Lifshitz black holes do no exhibit constant curvature invariants.

To prove that the black hole metric \eqref{LifthitzBH} is solution of ZDG, it is convenient to consider the following ansatz for $e_2$
\begin{equation}
\begin{split}
	e_2{}^{\hat{1}} &= \dfrac{ (\alpha -11\sigma)r^2 + 4M\ell^2\sigma }{2\beta r^2}  \left(\dfrac{r}{\ell}\right)^3 \sqrt{1 - \dfrac{M\ell^2}{r^2}} \, dt \, , \\
	e_2{}^{\hat{2}} &= \dfrac{ \alpha + 5\sigma }{ 2\beta }\, \dfrac{r}{\ell} \, dx \, , \\
	e_2{}^{\hat{3}} &= \dfrac{ \alpha - 7\sigma }{ 2\beta } \left( \dfrac{r^2}{\ell^2} - M \right)^{-1/2} dr \, .
\end{split}
	\label{e2LifshitzBH}
\end{equation}

The three-dimensional Lifshitz black hole of \cite{NMG:LifshitzBH} solves the ZDG equations of motion if $\mathcal{A}$ and $\beta$ satisfy the conditions (\ref{ZDG:LifshitzCond}) evaluated at $z=3$; namely
\begin{equation}
	\mathcal{A} = -\dfrac{ \big( \alpha (\alpha - 6\sigma ) - 55 \big)^2 ( \alpha - 13\sigma ) }{ 2\sigma ( \alpha - 7\sigma )^4 } \, , \quad
	\beta = - \dfrac{ \alpha ( \alpha - 6\sigma ) - 55 }{2\sigma ( \alpha - 7\sigma ) } \,,
\label{ZDG:LifshitzBHCond}
\end{equation}
or, equivalently,
\begin{equation}
	\dfrac{\mathcal{A}}{\beta} = \dfrac{ \big( \alpha ( \alpha - 6\sigma) - 55 \big) \, \left( \alpha - 13\sigma \right) } { \big( \alpha - 7\sigma \big)^3 } .
\end{equation}
Here we are assuming $\alpha \neq 11\sigma$, $\alpha \neq 7\sigma$, and $\alpha \neq -5\sigma$ for \eqref{e2LifshitzBH}-\eqref{ZDG:LifshitzBHCond} to be well defined.

We can see that the `metric' associated with the second dreibein \eqref{e2LifshitzBH} can be re-written as a global factor times an asymptotically $z=3$ Lifshitz black hole. Actually, the line element of $g_2$ reads
\begin{equation}
	ds_2{}^2 = \Omega^2 \left\lbrace -\left( \dfrac{r^2}{\ell^2} - M \right) \left( \dfrac{r^2}{\ell^2} + \mu \right)^2 d\tilde{t}^{\:2} + \dfrac{r^2}{\ell^2} d\tilde{x}^2 +\dfrac{dr^2}{\left( \dfrac{r^2}{\ell^2} - M \right)} \right\rbrace \,,
	\label{g2LifshitzBH}
\end{equation}
where we (re-)defined
\begin{equation}
	\Omega^2 = \dfrac{ (\alpha - 7\sigma)^2 }{4\beta^2} \,, \quad \mu = \dfrac{4M\sigma}{\alpha - 11\sigma} \,, \quad
	\tilde{t} = \dfrac{ \alpha - 11\sigma }{ \alpha - 7\sigma } \,,
	 \quad \tilde{x} = \dfrac{ \alpha - 5\sigma }{ \alpha - 7\sigma } \,.
\end{equation}
The parameter $\mu$ is positive provided $sign(M\sigma)(\alpha - 11\sigma) > 0$.

The `spacetime' described by the line element \eqref{g2LifshitzBH} has non-constant curvature invariants with the same divergences structure than \eqref{LifthitzBH}. However, it is important to remark that black hole \eqref{g2LifshitzBH} is different from the one found in \cite{NMG:LifshitzBH}, representing a one-parameter deformation of the latter.  The presence of the parameter $\mu$ makes the black hole \eqref{g2LifshitzBH} not locally equivalent to \eqref{LifthitzBH}, and it may also contribute to global properties of the geometry like the value of its mass. Both black holes share, however, the same asymptotic behaviour, approaching $z=3$ Lifshitz space at large $r$.

Notice that, provided both $M$ and $\mu$ are positive, black hole \eqref{g2LifshitzBH} has positive temperature, given by
\begin{equation*}
	T = \dfrac{\sqrt{M}(M+\mu)}{2\pi\ell} \,.
\end{equation*}

It remains an open problem how to define conserved charges in this setup, in order to study, for example, the thermodynamics of the new black hole \eqref{g2LifshitzBH}.

\subsection{Schr\"{o}dinger Spacetimes in ZDG}
\label{SubSec:ZDGSchroedinger}
Now, let us consider other anisotropic spaces, the Schr\"{o}dinger spacetimes. These are a special case of the AdS-waves studied in \cite{ZDG:LogAdSWaves}, for which the wave profile functions \eqref{AdSWaveProfile} takes the values $f_+(u) =0$, $f_-(u) = f_0$, and $n_- = -2\nu$. The dreibein $e_1$ in this case is given by 
\begin{equation*}
	e_1{}^{\hat{1}} = \dfrac{\ell}{y} \left(\dfrac{\ell^{\nu}}{y^{\nu}} du + \dfrac{y^{\nu}}{\ell^{\nu}} dv \right) \,, \quad
	e_1{}^{\hat{2}} = \dfrac{\ell}{y} \dfrac{y^{\nu}}{\ell^{\nu}} dv \,, \quad
	e_1{}^{\hat{3}} = \dfrac{\ell}{y} dy \: ,
\end{equation*}
while $e_2$ is given by
\begin{equation*}
\begin{split}
	e_2{}^{\hat{1}} &= \dfrac{\ell}{y} \left[ \left(\gamma - \dfrac{2\sigma\nu(1+\nu)}{\beta} \right)\dfrac{\ell^{\nu}}{y^{\nu}} du + \gamma\, \dfrac{y^{\nu}}{\ell^{\nu}} dv \right] \: , \\
	e_2{}^{\hat{2}} &= \dfrac{\ell}{y} \left[ - \dfrac{2\sigma(1+\nu)}{\beta} \dfrac{\ell^{\nu}}{y^{\nu}} du +  \gamma\, \dfrac{y^{\nu}}{\ell^{\nu}} dv \right] \: , \\
	e_2{}^{\hat{3}} &= \dfrac{\ell\gamma}{y} dy \: .
\end{split}
\end{equation*}
For this geometry to be a solution of ZDG, the parameters of the theory must satisfy the following relations
\begin{equation}
	\alpha = -\sigma \big( 2\beta +1 - 8\nu -8\nu^2 \big) \,, \quad
	\mathcal{A} = -\dfrac{\beta^{\: 2}(\beta - 1)}{\big( \beta - 4\nu(1+\nu) \big)^2} \,, \quad
	\gamma = -\dfrac{\sigma \big( \beta-4\nu(1+\nu) \big)}{\beta} \,. \nonumber
\end{equation}

Since the Schrödinger spaces are particular cases of the AdS waves, the spin-two field $g_2$ is a global rescaling of $g_1$ as we discussed in section \ref{SubSec:SelfDuality}. To avoid repetition we address the reader's attention to ref. \cite{ZDG:LogAdSWaves} for details.

\subsection{Warped Anti-de Sitter Spacetimes in ZDG}
\label{SubSec:ZDGWAdS}
We already mentioned that a special case of Schr\"{o}dinger spaces is the so-called null WAdS geometry. This is therefore a solution of ZDG as well. Besides, the theory admits also spacelike and timelike WAdS spaces. Interestingly, the spacelike streched WAdS as well as the black hole that is locally equivalent (and asymptote) to it, namely \eqref{WAdSADM} solves the ZDG equations of motion \eqref{R1eom}-\eqref{R2eom}. So, the first dreibein has the form
\begin{equation}
\begin{split}
	e_1{}^{\hat{1}} &= dt + \left( \nu r - \frac12 \sqrt{(\nu^2+3)r_+ r_-} \right) d\varphi \: , \\
	e_1{}^{\hat{2}} &= \frac12 \sqrt{(\nu^2+3)(r-r_+)(r-r_-)} d\varphi \: , \\
	e_1{}^{\hat{3}} &= \dfrac{\ell}{\sqrt{(\nu^2+3)(r-r_+)(r- r_-)}} dr \:,
\end{split}
\end{equation}
while the second dreibein has the form
\begin{equation}
\begin{split}
	e_2{}^{\hat{1}} &= \dfrac{\alpha - \sigma(4\nu^2-3)}{4\beta} \left[ 2 dt + \left( 2\nu r - \sqrt{(\nu^2+3)r_+ r_-} \right) d\varphi \right] \: , \\
	e_2{}^{\hat{2}} &= \dfrac{\alpha + \sigma(2\nu^2-3)}{4\beta} \sqrt{(\nu^2+3)(r-r_+)(r-r_-)} d\varphi \: , \\
	e_2{}^{\hat{3}} &= \dfrac{\ell}{2\beta}\dfrac{\alpha + \sigma(2\nu^2-3)}{\sqrt{(\nu^2+3)(r-r_+)(r-r_-)}} dr \:.
	\label{e2WAdSBH}
\end{split}
\end{equation}

For these black hole geometries to be solutions of ZDG, the parameters $\alpha, \mathcal{A}$, and $\beta$ must satisfy
\begin{equation}
\begin{split}
	\dfrac{\mathcal{A}}{\beta} &= \dfrac{ \left( \alpha - \sigma(4\nu^2-3) \right) }{ \left( \alpha + \sigma(2\nu^2-3) \right)^6} \left( 20\nu^4 - 12\sigma (\alpha + \sigma) \nu^2 + (\alpha - 3\sigma)^2 \right) \\
	& \times \left( 8\sigma\nu^6 + 12(\alpha - 3\sigma)\nu^4 - 12\sigma(\alpha^2-3)\nu^2 + (\alpha-3\sigma)^3 \right) \,.
\end{split}
\end{equation}

The `metric' associated to the second dreibein of the spacelike WAdS black hole \eqref{e2WAdSBH} can be put in the following form
\begin{equation}
\begin{split}
	ds_2{}^2 &= \Omega^2 \left\lbrace d\tilde{t}^{\:2} + 2\left(\nu r - \dfrac{1}{2}\sqrt{(\nu^2+3)r_+r_-} \right) d\tilde{t}d\tilde{\varphi} + \dfrac{\ell^2 dr^2}{(\nu^2+3)(r-r_+)(r-r_-)} \right. \\
	& \left. + \dfrac{r}{4}\left( 3(\nu^2-1)r + (\nu^2+3)(r_+ + r_-) -4\nu\sqrt{(\nu^2+3)r_+r_-} \right) d\tilde{\varphi}^2  \right\rbrace
	\label{g2WAdSBH}
\end{split}
\end{equation}
where we (re-)defined
\begin{equation}
\begin{split}
	\Omega^2 = \dfrac{ 9(\nu^2-1)^2 }{4\beta^2} \,, \quad \alpha = \nu^2\sigma \,, 
	\quad \tilde{t} = - t \,, 
	\quad \tilde{\varphi} = -\varphi \,,
	 \label{WAdSBHscalings}
\end{split}
\end{equation}
and where the coupling constants take the values
\begin{equation}
	\mathcal{A} = -\dfrac{(\nu^2-3)}{2} \,, \quad \beta = \dfrac{3(\nu^2-1)}{2} \,. \nonumber
\end{equation}
It is worth mentioning that the `spacetime' described by \eqref{g2WAdSBH} does not differ globally from \eqref{WAdSADM} since the parameter rescaling given by \eqref{WAdSBHscalings} does not introduce any angular deficit.

\section{Conclusions}
\label{Sec:Conclusions}

In this paper we have constructed explicit solutions of the recently proposed Zwei-Dreibein Gravity (ZDG) theory, which is a consistent bigravity theory in three dimensions. Among the exact solutions of this theory we have found geometries with anisotropic scale invariance with and without Galilean invariance. These are three-dimensional analogues of the Schr\"{o}dinger and Lifshitz spaces, respectively, which were conjectured to be dual to strongly coupled condensed matter systems. We have also found static circularly symmetric black holes that asymptote Lifshitz space, which is remarkable as they represent non-constant curvature solutions of New Massive Gravity that persist after the introduction of the infinite set of higher-curvature terms of ZDG. Warped Anti-de Sitter black holes were also derived. For the Lifshitz, Schr\"{o}dinger and Warped Anti-de Sitter spacetimes, the two spin-two fields of the theory are related each other by a global rescaling and a coordinate transformation. For the Lifshitz black hole, in contrast, the second `metric' represents a new Lifshitz black hole that asymptote $z=3$ Lifshitz geometry, representing a new family of one-parameter Lifshitz black holes. Whether the two dynamical fields of the theory have a clear geometric interpretation remains as an open problem. In addition, it would be interesting to find out whether ZDG contains in its spectrum new anisotropic scale invariance solutions, like Lifshitz black holes with $z \neq 3$. An open problem is how to define conserved charges for this theory; finding a method to do so would be important in order to study, for example, the thermodynamics of the bigravity black holes. The results of this paper give additional motivations to study three-dimensional massive (bi)gravity theories as toy models to study AdS/CFT holography and its ramifications.

%\appendix
%\section{Some title}
%Please always give a title also for appendices.

\acknowledgments

The work of A.F.G. is supported by Consejo Nacional de Investigaciones Cient\'ificas y T\'ecnicas (CONICET) and by Universidad de Buenos Aires (UBA). A.F.G. thanks G. Giribet and W. Merbis for useful comments and remarks on the manuscript, and to E. Bergshoeff and J. Rossell for discussions on Zwei-Dreibein Gravity. A.F.G. would like to thank the hospitality of Centre for Theoretical Physics of the University of Groningen where part of this work was done.


\end{document}